\documentstyle[12pt]{article}
\newcommand{\be}{\begin{equation}}
\newcommand{\ee}{\end{equation}}

\newcommand{\al}{\alpha}

\newcommand{\bet}{\beta}

\newcommand{\Om}{\Omega}
\newcommand{\om}{\omega}

\newcommand{\bzi}{{\bar{z}}}

\newcommand{\bz}{\bar{z}}
\newcommand{\by}{\bar{y}}
\newcommand{\G}{\Gamma}

\newcommand{\La}{\Lambda}

\newcommand{\M}{{\cal M}}
\newcommand{\fm}{{\cal F}({\cal M})}
\newcommand{\Pen}{{\cal P}}     
\newcommand{\Pn}{\tilde{\cal P}_0}     
\newcommand{\pn}{\tilde{\cal P}}

\newcommand{\lo}{L_+^\uparrow}
\newcommand{\la}{\lambda}

\newcommand{\dz}{\wedge}

\newcommand{\ba}{\begin{array}}
\newcommand{\ea}{\end{array}}
\newcommand{\beq}{\begin{eqnarray}}
\newcommand{\eeq}{\end{eqnarray}}
\newtheorem{lm}{Lemma}
\newtheorem{th}{Theorem}
\newtheorem{pr}{Proposition}
\newtheorem{co}{Corollary}
\newtheorem{rem}{Remark}
\newtheorem{deff}{Definition}
\newcommand{\bd}{\begin{deff}}
\newcommand{\ed}{\end{deff}}
\newcommand{\lij}{\la^i_{~j}}
\newcommand{\bl}{\begin{lm}}
\newcommand{\el}{\end{lm}}
\newcommand{\bp}{\begin{pr}}
\newcommand{\ep}{\end{pr}}
\newcommand{\bt}{\begin{th}}
\newcommand{\et}{\end{th}}
\newcommand{\bc}{\begin{co}}
\newcommand{\ec}{\end{co}}
\newcommand{\brm}{\begin{rem}}
\newcommand{\erm}{\end{rem}}

\newcommand{\der}{\mbox{d}}
\begin{document}

\thispagestyle{empty}

\title {ELLIPTIC FIBRATIONS ASSOCIATED WITH THE EINSTEIN SPACETIMES
\footnote{Research supported 
by 
Komitet Bada\'{n} Naukowych (Grant nr 2 P03B 017 12), Consorzio per 
lo Sviluppo Internazionale dell Universita degli Studi di Trieste and the 
Erwin Schr\"{o}dinger International Institute for Mathematical Physics.}\\ 
\vskip 1.truecm
{\small {\sc Pawe\l  ~Nurowski}}\\
{\small {\it Dipartimento di Scienze Matematiche }}\\
\vskip -0.3truecm
{\small {\it Universita degli Studi di Trieste, Trieste, Italy}}\\
\vskip -0.3truecm
{\small {\it and}}\\
\vskip -0.3truecm
{\small {\it Department of Mathematics}}\\
\vskip -0.3truecm
{\small {\it King's College, London, UK}}
\footnote{Permanent address: 
{\small{\small {\it 
Katedra Metod Matematycznych Fizyki, Wydzia\l ~Fizyki, 
Uniwersytet Warszawski, ul. 
Ho\.za 74, Warszawa, Poland, e-mail: nurowski@fuw.edu.pl
}}}}}
\author{\mbox{}}
\maketitle
\begin{abstract}
Given a conformally nonflat Einstein spacetime we define a fibration 
$\pn$ over it. The fibres of this fibration are elliptic curves 
(2-dimensional tori) or their degenerate counterparts. Their topology 
depends on the algebraic type of the Weyl tensor of the 
Einstein metric.\\
The fibration $\pn$ is a double branched cover of the bundle $\Pen$ of 
null direction over the spacetime and is equipped with six linearly 
independent 1-forms which satisfy certain relatively simple system 
of equations.
\end{abstract}

\newpage
\noindent
{\large {\bf 1.}} In Ref. \cite{bi:nur2} we defined a certain 
differential system $\cal I$ on an open set $U$ of 
${\bf R}^6$. We showed that a pair $(U, {\cal I})$ naturally 
defines a 4-dimensional conformally non-flat Lorentzian spacetime 
$({\cal M}, g)$ which satisfies the Einstein equations $R_{ij}=\la g_{ij}$. 
In this paper we prove the converse satement. In particular, we give 
a construction which associates 
a certain 6-dimensional elliptic fibration $\tilde{\cal P}$ with 
any conformally non-flat Lorentzian Einstein spacetime. Moreover, we show  
how $\tilde{\cal P}$ may be equipped with a unique differential system 
which has all the properties of the system $\cal I$ on $U$. \\

\noindent
We brieflly recall the definitions of the geometrical 
objects we need in the following. Let $\M$ be a 4-dimensional 
oriented and time-oriented 
manifold equipped with a Lorentzian metric $g$ of signature $(+,+,+,-)$. 
It is convenient to introduce a null frame $(m,~\bar{m},~k,~l)$ on $\M$ 
with a coframe $\theta^i=(\theta^1,~\theta^2,~\theta^3,~\theta^4)=
(M,~\bar{M},~K,~L)$ so that 
\be 
g=g_{ij}\theta^i\theta^j=M\bar{M}-KL.
\footnote{Such expressions as $\theta^i\theta^j$ mean the symmetrized tensor 
product, e.g. $\theta^i\theta^j=\frac{1}{2}(\theta^i\otimes\theta^j+ 
\theta^j\otimes\theta^i)$. Also, we will denote by round (resp. square) 
brackets the symmetrization (resp. antisymmetrization) of indices, e.g. 
$a_{(ik)}=\frac{1}{2}(a_{ik}+a_{ki})$, $a_{[ik]}=
\frac{1}{2}(a_{ik}-a_{ki})$, etc.}
\label{eq:met} 
\ee 
The Lorentz group $L$ consists of matrices $\lij\in{\bf GL}(4,{\bf C})$ such that 
$$g_{jl}=g_{ik}\lij\la^k_{~l}$$ 
\be
\la^2_{~2}=\overline{\la^1_{~1}}\quad\quad \la^2_{~1}=\overline{\la^1_{~2}}
\quad\quad \la^2_{~3}=\overline{\la^1_{~3}}\quad\quad \la^2_{~4}=
\overline{\la^1_{~4}}
\ee
$$
\la^3_{~2}=\overline{\la^3_{~1}}\quad\quad\la^3_{~3}=
\overline{\la^3_{~3}}\quad\quad
\la^3_{~4}=\overline{\la^3_{~4}}\quad\quad
\la^4_{~2}=\overline{\la^4_{~1}}\quad\quad\la^4_{~3}=
\overline{\la^4_{~3}}\quad\quad
\la^4_{~4}=\overline{\la^4_{~4}}.
$$
We will denote the inverse of the Lorentz matrix $\lij$ by 
$\tilde{\la}^i_{~j}$.\\ 
The connected component of the identity element of $L$ is the proper 
ortochroneous Lorentz group, which we denote by $\lo$.\\
Given $g$ and $\theta^i$ the connection 1-forms 
$\G_{ij}=g_{ik}\G^k_{~j}$ are uniquely defined by 
\be
\der\theta^i=-\G^i_{~j}\dz\theta^j,~~~~~~~\G_{ij}+\G_{ji}=0.\label{eq:kon}
\ee
The connection coefficients $\G_{ijk}$ are determined 
by $\G_{ij}=\G_{ijk}\theta^k$.\footnote{We lower and raise indices by means 
of the metric and its inverse} 
Using them we define the curvature 2-forms ${\cal R}^k_{~i}$, the 
Riemann tensor $R^{i}_{~jkl}$, the Ricci tensor 
$R_{ij}$ and its scalar $R$ by 
$$
{\cal R}^k_{~i}=\frac{1}{2}R^k_{~imj}\theta^m\dz\theta^j=\der\G^k_{~i}+
\G^k_{~j}\dz\G^j_{~i},~~~R_{ij}=R^k_{~ikj},~~~R=g^{ij}R_{ij}.
$$
We also introduce the traceless Ricci tensor by 
$$
S_{ij}=R_{ij}-\frac{1}{4}g_{ij}R.
$$
Note that the vanishing of $S_{ij}$ is equivalent to the Einstein equations 
$R_{ij}=\la g_{ij}$ for the metric $g$. We define the Weyl tensor 
$C^i_{~jkl}$ by
$$
C_{ijkl}=R_{ijkl}+\frac{1}{3}Rg_{i[k}g_{l]j}+R_{j[k}g_{l]i}+R_{i[l}g_{k]j},
$$
and its spinorial coefficients $\Psi_\mu$ by
\footnote{Theses equations are coppied from Ref. \cite{bi:nur1}. 
There, the curvature 2-forms ${\cal R}_{ij}$ 
were denoted by  $\Om_{[ij]}$. 
We take this opportunity to mention that a sign missprint was present 
in the expressions for $\Om_{[23]}$ and $\Om_{[13]}$ in appendix B of 
\cite{bi:nur1}. The correct 
formulae (B1) and (B2) of \cite{bi:nur1} should have minus sign in front of 
$\frac{1}{12}R$ in the expressions for 
$\Om_{[23]}$ and $\Om_{[13]}$. Note also 
that because of the orientation change, formulae of \cite{bi:nur1} 
translate to the present paper by respecting the following rules. 
(a) Put $N$ and $P$ of \cite{bi:nur1} to be equal to 
the present $L$ and $\bar{M}$, respectively. 
(b) Replace all the primes in \cite{bi:nur1} by bars, 
e.g. $\Psi_2'\to\bar{\Psi}_2$, $z'\to\bar{z}$, etc.  
(c) Interchange the indices 3 and 4 in all of the formulae.}
\beq                 
&{\cal R}_{23}=\bar{\Psi}_4 \bar{M}\dz K+\bar{\Psi}_3 (L\dz K-M\dz \bar{M})+
(\bar{\Psi}_2 +\frac{1}{12}R)L\dz M\nonumber\\
&+\frac{1}{2}S_{33}M\dz K+\frac{1}{2}S_{32}(L\dz K+M\dz\bar{M})+
\frac{1}{2}S_{22}L\dz\bar{M},\nonumber\\
&\quad\nonumber\\
&{\cal R}_{14}=(-\bar{\Psi}_2-\frac{1}{12}R)\bar{M}\dz K-
\bar{\Psi}_1(L\dz K-M\dz \bar{M})-
\bar{\Psi}_0 L\dz M\nonumber\\
&-\frac{1}{2}S_{11}M\dz K-\frac{1}{2}S_{41}(L\dz K+M\dz\bar{M})-
\frac{1}{2}S_{44}L\dz\bar{M},\nonumber\\
&\quad\nonumber\\
&\frac{1}{2}({\cal R}_{43}-{\cal R}_{12})=\bar{\Psi}_3 \bar{M}\dz K+
(\bar{\Psi}_2-\frac{1}{24}R) 
(L\dz K-M\dz \bar{M})+\bar{\Psi}_1 L\dz M\nonumber\\
&+\frac{1}{2}S_{31}M\dz K+\frac{1}{4}(S_{12}+S_{34})(L\dz K+M\dz\bar{M})+
\frac{1}{2}S_{42}L\dz\bar{M}.\nonumber
\eeq~\\

\noindent
{\large {\bf 2.}} Let $\fm$ denote the bundle of oriented and time oriented null 
coframes over $\cal M$. This means that $\fm$ is the set of all 
equally oriented 
null coframes $\theta^i$ at all points of $\cal M$. The mapping 
$\pi:\fm\to\cal M$, which maps a coframe $\theta^i$ at $x\in\cal M$ onto 
$x$, gives the canonical projection. A fibre $\pi^{-1}(x)$
in $\fm$ consists of all the null coframes at point $x$ which have the same 
orientation and time orientation. If $\theta^j$ is a null coframe at 
$x\in\cal M$ then any other equivalently oriented null coframe 
at $x$ is given by $\theta^{'i}=\lij\theta^j$, where $\lij$ is a 
certain element of $\lo$. This defines an action of $\lo$ on $\fm$.  
Thus, $\fm$ is a 10-dimensional principal fibre bundle  with 
$\lo$ as its structural group.\\ 
It is well known that the bundle $\fm$ is equipped with a natural 
4-covector-valued 1-form $e^i$, $i=1,2,3,4$, the Cartan soldering 
form, which is defined as follows. Take any vector $v_c$ tangent to $\fm$ 
at a point $c$. Let $c$ be in the fibre 
$\pi^{-1}(x)$ over a point $x\in\cal M$. 
This means that $c$ may be identified with a certain null coframe 
$\theta^i_c$ at $x$. Then, the formal 
definition of $e^i$ reads: $e^i(v_c)=\theta^i_c(\pi_*v_c)$. 
The first two components of $e^i$ are complex and mutually conjugated. 
The remaining two are real. Altogether, they constitute a system of 
four well defined linearly independent 1-forms on $\fm$. In the following 
we will denote them by 
\be
F=e^1=\overline{e^2}\quad\quad\quad T=\bar{T}=e^3\quad\quad\quad
\La=\bar{\La}=e^4.
\label{eq:fbb}
\ee
A theorem which we present below is a null coframe reformulation of 
the Elie Cartan theorem on affine conections. 
\bt
Let $e^i=(F,\bar{F},T,\La)$ be the soldering form on $\fm$. Then 
\begin{itemize}
\vskip -1.truecm
\item[(i)] the system of equations 
\be
\der e^i +\omega^i_{~j}\dz e^j=0,\quad\quad 
g_{ki}\omega^i_{~j}+g_{ji}\omega^i_{~k}=0
\label{eq:cse}
\ee
for a matrix of complex-valued 1-forms $\omega^i_{~j}$ $(i,j=1,2,3,4)$ on 
$\fm$ has unique solution, 
\item[(ii)] $\omega^i_{~j}$ uniquely defines six complex-valued 1-forms 
$(E,\bar{E},\G,\bar{\G},\Om,\bar{\Om})$ by  
\be
\omega^i_{~j} =
\left(\begin{array}{cccc}
\bar{\Om}-\Om&0&-E&-\G\\
0&\Om-\bar{\Om}&-\bar{E}&-\bar{\G}\\
-\bar{\G}&-\G&\Om+\bar{\Om}&0\\
-\bar{E}&-E&0&-\Om-\bar{\Om}
\end{array}\right),\label{eq:cso}
\ee
\item[(iii)] the forms 
$(F,\bar{F},T,\La,E,\bar{E},\G,\bar{\G},\Om,\bar{\Om})$ 
are linearly independent at each point of $\fm$.
\end{itemize}
\et
For completness we sketch the proof.\\

\noindent
First, we show that if there is a solution to (\ref{eq:cse}) then it 
is unique. To do this we assume an existence of two solutions 
$\omega^i_{~j}$ and $\hat{\omega}^i_{~j}$.\\
Subtracting $\der e^i +\hat{\omega}^i_{~j}\dz e^j=0$ from 
$\der e^i +\omega^i_{~j}\dz e^j=0$ we get 
\be
(\omega^i_{~j}-\hat{\omega}^i_{~j})\dz e^j=0. \label{eq:d1}
\ee
Now, let $e^\mu$, 
$\mu=1,2,3,4,5,6$, be a system of 1-forms such that the ten 1-forms 
$(e^i,e^\mu)$ constitute a basis of 1-forms on $\fm$. Let 
$\omega^i_{~j}=\omega^i_{~jk}e^k+\omega^i_{~j\mu}e^\mu$ and 
$\hat{\omega}^i_{~j}=\hat{\omega}^i_{~jk}e^k+\hat{\omega}^i_{~j\mu}e^\mu$ 
be the corresponding decompositions of the solutions. Then (\ref{eq:d1}) 
easilly yields $\omega^i_{~j\mu}=\hat{\omega}^i_{~j\mu}$ and 
$\omega_{i[jk]}=\hat{\omega}_{i[jk]}$. The defining properties 
of the solutions give also $\omega_{(ij)k}=0=\hat{\omega}_{(ij)k}$. Now,   
due to the identity $A_{ijk}=A_{i[jk]}-A_{j[ik]}-A_{k[ij]}$, which is true 
for any $A_{ijk}$ such that $A_{(ij)k}=0$, we get  
$\omega_{ijk}=\hat{\omega}_{ijk}$. This shows that 
$\omega^i_{~j}=\hat{\omega}^i_{~j}$, hence the uniqueness.\\

\noindent 
We pass to the construction of a solution.\\
Given a sufficiently small neighbourhood $\cal O$ in $\cal M$ we 
identify $\pi^{-1}({\cal O})$ with ${\cal O}\times\lo$. Then, 
the soldering form may be written as 
\be
e^i=\lij\theta^j.\label{eq:fcc}
\ee
Taking $\der e^i$ and using the equations (\ref{eq:kon}) 
one easilly finds that 
\be
\om^i_{~j}=\la^i_{~k}\G^k_{~m}\tilde{\la}^m_{~j}-\der\la^i_{~k}
\tilde{\la}^k_{~j}
\label{eq:roz}
\ee
is a solution to (\ref{eq:cse}). Given this solution one defines 
the forms $(E,\G,\Om)$ by 
\be
E=-\om^1_{~3}\quad\quad\quad\G=-\om^1_{~4}\quad\quad\quad
\Om=\frac{1}{2}(\om^2_{~2}+\om^3_{~3}). \label{eq:fbc}
\ee
This is in accordance with (\ref{eq:cso}) due to $\omega_{(ij)}=0$ and the 
reality properties of $e^i$.\\ 
                   
\noindent
To prove the linear independence of the system 
$(F,\bar{F},T,\La,E,\bar{E},\G,\bar{\G},\Om,\bar{\Om})$ 
it is enough to observe that the six 1-forms 
$\der\la^1_{~k}\tilde{\la}^k_{~3}$,  
$\der\la^2_{~k}\tilde{\la}^k_{~3}$, $\der\la^1_{~k}\tilde{\la}^k_{~4}$,     
$\der\la^2_{~k}\tilde{\la}^k_{~4}$, $\der\la^1_{~k}\tilde{\la}^k_{~1}$ and 
$\der\la^3_{~k}\tilde{\la}^k_{~3}$ constitute a basis of right  
invariant forms on $\lo$.\\

\noindent
The theorem is proven.\\

\noindent
{\large {\bf 3.}} Consider the ten well defined 
forms $(F,\bar{F},T,\La,E,\bar{E},\Om,\bar{\Om},\G,\bar{\G})$ on $\fm$ 
given by (\ref{eq:fbb}) and (\ref{eq:fbc}). 
The differentials of the first four forms are given by (\ref{eq:cse}) and 
(\ref{eq:cso}). In the basis 
$(F,\bar{F},T,\La,E,\bar{E},\Om,\bar{\Om},\G,\bar{\G})$ they assume the form 
\beq
&{\rm d}F=(\Om -\bar{\Om})\dz F+E\dz T+\G\dz\La,\label{eq:f1}\\
&{\rm d}T=\bar{\G}\dz F+\G\dz \bar{F}-(\Om +\bar{\Om})\dz T,\label{eq:k1}\\
&{\rm d}\La=\bar{E}\dz F+E\dz\bar{F}+ (\Omega+\bar{\Omega})\dz\La. 
\label{eq:l1}
\eeq
The differentials of the other forms may be easilly calculated using the 
local representation (\ref{eq:roz}) and the well known structure 
equation 
\be
\der\om^i_{~j}=-\om^i_{~k}\dz\om^k_{~j}+\frac{1}{2}R^k_{~smn}\la^i_{~k}
\tilde{\la}^s_{~j}\tilde{\la}^m_{~l}\tilde{\la}^n_{~p}e^l\dz e^p.
\label{eq:fcd}
\ee
These diferentials are by far 
much more complicated than the differentials of $\der F$, $\der T$ and 
$\der\La$. In particular, the decompositions of 
$\der F$, $\der T$ and $\der\La$ onto the basis of 2-forms associated with 
$(F,\bar{F},T,\La,E,\bar{E},\Om,\bar{\Om},\G,\bar{\G})$ have only constant 
coefficients. It turns out that in 
the differentials $\der E$, $\der \Om$ and $\der\G$  
coefficients which are functions appear. The zero sets of these functions 
have a well defined geometrical meaning and define certain subsets of $\fm$. 
Now, the hope is that when we restrict to such subsets then 
the differentials of $E$, $\Om$, and $\G$ 
will have much simpler form than their differentials on the whole $\fm$. 
Our aim now is to study this possibility.\\

\noindent
{\large {\bf 4.}} We concentrate on the analysis of $\der E=-\der\om^1_{~3}$.\\
Let us introduce the matrices $\lij(w,z,y)$ and $\lij(w',z',y')$ such that 
$$
\lij(w,z,y)=
\left(\begin{array}{cccc}
|w|w^{-1}(1+\bar{y}\bar{z})&|w|w^{-1}\by z
&|w|w^{-1}(1+\bar{y}\bar{z})z&|w|w^{-1}\by\\
\mbox{}\\
w|w|^{-1}y\bz&w|w|^{-1}(1+yz)
&w|w|^{-1}(1+yz)\bz&w|w|^{-1}y\\
\mbox{}\\
|w|(1+\bar{y}\bar{z})y&|w|(1+yz)\by&|w||1+\bar{y}\bar{z}|^2&|w||y|^2\\
\mbox{}\\
\bz|w|^{-1}&z|w|^{-1}&|z|^2|w|^{-1}&|w|^{-1}
\end{array}\right),
$$
$$
\lij(w',z',y')=
\left(\begin{array}{cccc}
|w'|w'^{-1}\by' z'&|w'|w'^{-1}(1+\bar{y}'\bar{z}')
&|w'|w'^{-1}\by'&|w'|w'^{-1}(1+\bar{y}'\bar{z}')z'\\
\mbox{}\\
w'|w'|^{-1}(1+y'z')&w'|w'|^{-1}y'\bz'&w'|w'|^{-1}y'&w'|w'|^{-1}(1+y'z')\bz'\\
\mbox{}\\
|w'|(1+y'z')\by'&|w'|(1+\bar{y}'\bar{z}')
&|w'||y'|^2&y'|w'||1+\bar{y}'\bar{z}'|^2\\
\mbox{}\\
z'|w'|^{-1}&\bz'|w'|^{-1}&|w'|^{-1}&|z'|^2|w'|^{-1}
\end{array}\right). 
$$  
Then, it is well known that $\lo$ can be represented by 
\be
\lo={\cal U}\cup {\cal U}'
\ee
where
\be
{\cal U}=\{~\lij(w,z,y)~~~{\rm s.t.}~~~(w,z,y)\in{\bf C}^3, ~w\neq 0~\}
\ee
\be
{\cal U}'=\{~\lij(w',z',y')~~~{\rm s.t.}~~~(w',z',y')\in{\bf C}^3, ~w'\neq 0~\}. 
\ee
On the intersection ${\cal U}\cap {\cal U}'$, the coordinates $(w,z,y)$ and 
$(w',z',y')$ shall be related by 
\be
w=-\frac{w'}{z'^2}\quad\quad\quad 
z=\frac{1}{z'}\quad\quad\quad 
y=-z'(1+y'z')\label{eq:wzy}
\ee
\be
w'=-\frac{w}{z^2}\quad\quad\quad 
z'=\frac{1}{z}\quad\quad\quad 
y'=-z(1+yz).
\ee
Thus, we can cover any $\pi^{-1}({\cal O})\cong({\cal O}\times\lo)$ 
by the two charts 
${\cal O}\times{\cal U}$ and ${\cal O}\times{\cal U}'$.\\ 
Now, on $\cal O$ consider the coframe $\theta^i$ of (\ref{eq:met}).
Inserting $\lij=\lij(w,z,y)$ or $\lij(w',z',y')$ to the formulae 
(\ref{eq:fcc}), (\ref{eq:roz}), (\ref{eq:fcd}) and using the definitions 
(\ref{eq:met}), (\ref{eq:fbb}), (\ref{eq:fbc}) we easilly obtain the 
following two lemmas.
\bl
On ${\cal O}\times{\cal U}$ the forms 
$F$, $T$, $\La$, $E$, $\Om$ and $\G$ read 
\be
F=\frac{|w|}{w}[(1+\bar{y}\bar{z})(M+zK)+\bar{y}z\bar{M}+\bar{y}L],
\label{eq:ff1}
\ee
\be
T=|w|[|1+\bar{y}\bar{z}|^2K+y(1+\bar{y}\bar{z})M+
\bar{y}(1+yz)\bar{M}+y\bar{y}L],
\ee
\be
\La=\frac{1}{|w|}[L+z\bar{z}K+z\bar{M}+\bar{z}M]\label{eq:fl1}
\ee
\be
E=\frac{1}{w}[\der z+\G_{32}+z(\G_{21}+\G_{43})+z^2\G_{14}],\label{eq:eog1}
\ee
\be
\Om =-\frac{1}{2}\frac{{\rm d}{w}}{w}-y{\rm d}z-y\G_{32}-\frac{1}{2}(1+2yz)
(\G_{21}+\G_{43})-z(1+yz)\G_{14}\label{eq:eog2}
\ee
\be
\G =\bar{w} [{\rm d}\by-\by^2{\rm d}\bz-\by^2\G_{31}-\by(1+\by\bz)
(\G_{43}-\G_{21})-(1+\by\bz)^2\G_{24}].
\label{eq:eog3}
\ee                        
\el
\bl
On ${\cal O}\times{\cal U}'$ the forms 
$F$, $T$, $\La$, $E$, $\Om$ and $\G$ read\footnote{Note, that this Lemma 
follows from the previous one by applying transformations 
$M\leftrightarrow\bar{M}$, $K\leftrightarrow L$, $1\leftrightarrow 2$ and 
$3\leftrightarrow 4$, where the last two transformations reffer to the tetrad 
indices.}
\be
F=\frac{|w'|}{w'}[(1+\bar{y}'\bar{z}')(\bar{M}+z'L)+\bar{y}'z'M+\bar{y}'K],
\label{eq:ff1'}
\ee
\be
T=|w'|[|1+\bar{y}'\bar{z}'|^2L+y'(1+\bar{y}'\bar{z}')\bar{M}+
\bar{y}'(1+y'z')M+y'\bar{y}'K],
\ee
\be
\La=\frac{1}{|w'|}[K+z'\bar{z}'L+z'M+\bar{z}'\bar{M}]\label{eq:fl1'}
\ee
\be
E=\frac{1}{w'}[\der z'+\G_{41}+z'(\G_{12}+\G_{34})+z'^2\G_{23}],
\label{eq:eog1'}
\ee
\be
\Om =-\frac{1}{2}\frac{{\rm d}{w'}}{w'}-y'{\rm d}z'-y'\G_{41}-
\frac{1}{2}(1+2y'z')
(\G_{12}+\G_{34})-z'(1+y'z')\G_{23}\label{eq:eog2'}
\ee
\be
\G =\bar{w}' [{\rm d}\by'-\by'^2{\rm d}\bz'-\by'^2\G_{42}-\by'(1+\by'\bz')
(\G_{34}-\G_{12})-(1+\by'\bz')^2\G_{13}].
\label{eq:eog3'}
\ee                        
\el
Now one can easily find the differential of $\der E$. 
\bl
The decomposition of ${\rm d}E$ onto the basis 
$(F,\bar{F},T,\La,E,\bar{E},\Om,\bar{\Om},\G,\bar{\G})$ of 1-forms 
on $\pi^{-1}({\cal O})$ reads
$$
{\rm d}E=2\Om\dz E+\phi T\dz F+b (\La\dz T+F\dz\bar{F})+\bet F\dz\La
+\psi T\dz\bar{F}+a(\La\dz T-F\dz\bar{F})+\al F\dz\La,
$$
where $\phi,b,\bet,\psi,a,\al$ are well defined functions on 
$\pi^{-1}({\cal O})$.
\el
The functions $\phi,b,\bet,\psi,a,\al$ are given by 
$$
\phi=\frac{1}{|w|^2}\Phi,\quad\quad
b=\bar{w}(\frac{1}{2}\phi_\bzi +\by\phi),\quad\quad
\bet=\bar{w}^2(\frac{1}{2}\phi_{\bz\bz}+\by\phi_\bzi+\by^2\phi),
$$  
$$
\psi =\frac{1}{w^2}\Psi,\quad\quad
a=w(\frac{1}{4}\psi_z +y\psi),\quad\quad
\al=w^2(\frac{1}{12}\psi_{zz}+y^2\psi+\frac{1}{2}y\psi_z)+\frac{1}{12}R,
$$
$$
\Phi=
\frac{1}{2}S_{33} - \bar{z}S_{23} -
zS_{13}+ z\bar{z} (S_{12}+S_{34}) +\frac{1}{2} \bar{z}^2 S_{22}
+\frac{1}{2} z^2 S_{11} - \bar{z}^2 zS_{24} - z^2\bar{z}S_{14} + 
\frac{1}{2}z^2\bar{z}^2S_{44}
$$
and
$$
\Psi=\bar{\Psi}_4-4\bar{\Psi}_3z+6\bar{\Psi}_2z^2-4\bar{\Psi}_1z^3
+\bar{\Psi}_0z^4
$$
on ${\cal O}\times{\cal U}$ and by 
$$
\phi=\frac{1}{|w'|^2}\Phi',\quad\quad
b=\bar{w}'(\frac{1}{2}\phi_{\bz'} +\by'\phi),\quad\quad
\bet=\bar{w}'^2(\frac{1}{2}\phi_{\bz'\bz'}+\by'\phi_{\bz'}+\by'^2\phi),
$$  
$$
\psi =\frac{1}{w'^2}\Psi',\quad\quad a=w'(\frac{1}{4}\psi_{z'} +y'\psi),
\quad\quad\al=w'^2(\frac{1}{12}\psi_{z'z'}+y'^2\psi+\frac{1}{2}y'\psi_{z'})
+\frac{1}{12}R,
$$
$$
\Phi'=
\frac{1}{2}S_{44} - \bar{z}'S_{14} -
z'S_{24}+ z'\bar{z}' (S_{12}+S_{34}) +\frac{1}{2} \bar{z}'^2 S_{11}
+\frac{1}{2} z'^2 S_{22} - \bar{z}'^2 z'S_{13} - z'^2\bar{z}'S_{23} + 
\frac{1}{2}z'^2\bar{z}'^2S_{33}
$$
and 
$$
\Psi'=\bar{\Psi}_0-4\bar{\Psi}_1z'+6\bar{\Psi}_2z'^2-4\bar{\Psi}_3z'^3
+\bar{\Psi}_4z'^4
$$
on ${\cal O}\times{\cal U}'$.\\

\noindent
The following three cases are of particular interest.
\begin{itemize}
\item[({\bf A})] The metric $g$ of the 4-manifold $\cal M$ satisfies 
the Einstein equations $R_{ij}=\la g_{ij}$ and is not conformally flat. 
This case is characterized by $\Phi\equiv 0$ and $\Psi\not\equiv 0$.
\item[({\bf B})] The metric $g$ is conformally flat but not Einstein. 
This case corresponds to $\Psi\equiv 0$, $\Phi\not\equiv 0$.
\item[({\bf C})] The metric $g$ is of constant curvature. This means that 
$\Psi\equiv\Phi\equiv 0$.
\end{itemize}
In the first two cases there is a canonical choice of certain   
6-dimensional subsets in $\fm$. This is defined by the demand that 
on such sets certain components of $\der E$ should identically vanish.  
This approach is impossible in the case ({\bf C}) since this 
implies an immediate reduction of $\der E$ to the form 
\be
\der E=2\Om\dz E+\frac{1}{12}R\La\dz F.
\ee

\noindent
{\large {\bf 5.}} From now on we consider the case ({\bf A}) of the preceding section. 
This is the most interesting generic Einstein case.\\
Imposing the restrictions ({\bf A}) on $\der E$ we immediately see that  
$$
{\rm d}E=2\Om\dz E+
\psi T\dz\bar{F}+a (\La\dz T-F\dz\bar{F})+\al F\dz\La,
$$
where $\psi,a,\al$ are the same as in the Lemma 1. 
Since we are in the not conformally flat case ({\bf A}) we have 
$\psi\neq 0$. This makes possible the restriction to such 
a set ${\cal W}\subset\pi^{-1}({\cal O})$ in which $a$ identically vanish. 
Thus we consider   
\be
{\cal W}={\cal W}_1\cup{\cal W}_2,
\ee
where
$$
{\cal W}_1=\{~(x;w,z,y)\in({\cal O}\times{\cal U})~~~{\rm s.t.}~~~
\psi y+\frac{1}{4}\psi_z=0~\},
$$
$$
{\cal W}_2=\{~(x;w',z',y')\in({\cal O}\times{\cal U}')~~~{\rm s.t.}~~~
\psi y'+\frac{1}{4}\psi_{z'}=0~\},
$$
or (what is the same due to the nonvanishing of 
$w$ and $w'$) by 
\be
{\cal W}_1=\{~(x;w,z,y)\in({\cal O}\times{\cal U})~~~{\rm s.t.}~~~
\Psi y+\frac{1}{4}\Psi_z=0~\},
\ee
\be
{\cal W}_2=\{~(x;w',z',y')\in({\cal O}\times{\cal U}')~~~{\rm s.t.}~~~
\Psi' y'+\frac{1}{4}\Psi'_{z'}=0~\}.
\ee
On $\cal W$ we have   
$$
{\rm d}E=2\Om\dz E+\psi T\dz\bar{F}+\al F\dz\La.
$$
One can still simplify this relation by restricting to a 
subset $\Pn$ of ${\cal W}$ in which $\psi=-1$. Then, $\Pn$ is a subset of 
$\pi^{-1}({\cal O})$ which is given by
\be
\Pn=\pn_{01}\cup\pn_{02},
\ee
where 
\be
\pn_{01}=\{~(x;w,z,y)\in({\cal O}\times{\cal U})~~~{\rm s.t.}~~~
\Psi y+\frac{1}{4}\Psi_z=0, ~~~\Psi+w^2=0~\}
\ee
and
\be
\pn_{02}=
\{~(x;w',z',y')\in({\cal O}\times{\cal U}')~~~{\rm s.t.}~~~
\Psi' y'+\frac{1}{4}\Psi'_{z'}=0,~~~ \Psi'+w'^2=0~\}.
\ee
It follows from the construction that on $\Pn$ we have 
$$
{\rm d}E=2\Om\dz E+\bar{F}\dz T+\al F\dz\La.
$$

\noindent
{\large {\bf 6.}} We study the geometry and topology of the set $\Pn$.\\ 

\noindent
The equations defining $\Pn$ may be written as
\be
y=-\frac{1}{4}\frac{\Psi_z}{\Psi}, ~~~
y'=-\frac{1}{4}\frac{\Psi'_{z'}}{\Psi'}
\ee
\be
\Psi+w^2=0, ~~~\Psi'+w'^2=0.
\ee
We see that the first pair of equations eniquely 
subordinates $y$ to $z$ and $y'$ to $z'$. The second pair 
gives a realtion between $w$ and $z$ and $w'$ and $z'$. Thus, locally among 
the parameters $(x;w,z,y)$ in ${\cal O}\times{\cal U}$ (respectively, 
$(x;w',z',y')$ in ${\cal O}\times{\cal U}'$), only $x$ and $z$ 
(respectively, $x$ and $z'$) are free. This shows that $\Pn$ is 
6-dimensional. Moreover, $\Pn$ is fibred over $\cal O$ with 2-dimensional 
fibres. These are locally parametrized by $z$ or $z'$. To discuss the 
topology of fibres one observes that the relation between $w$ and $z$ 
(respectively, $w'$ and $z'$) is purely polynomial. Over every point of 
$\cal O$ it has the form 
\be
w^2=
-\bar{\Psi}_4+4\bar{\Psi}_3z-6\bar{\Psi}_2z^2+4\bar{\Psi}_1z^3
-\bar{\Psi}_0z^4\label{eq:torus} 
\ee
or  
\be
w'^2=
-\bar{\Psi}_0+4\bar{\Psi}_1z'-6\bar{\Psi}_2z'^2+4\bar{\Psi}_3z'^3
-\bar{\Psi}_4z'^4.\label{eq:torus'} 
\ee
Assume for a moment ({\bf a}) that $w=0$ and $w'=0$ are the allowed values of 
the parameters and ({\bf b}) that the equations 
\be
-\bar{\Psi}_4+4\bar{\Psi}_3z-6\bar{\Psi}_2z^2+4\bar{\Psi}_1z^3
-\bar{\Psi}_0z^4=0,\label{eq:tor}
\ee
\be
-\bar{\Psi}_0+4\bar{\Psi}_1z'-6\bar{\Psi}_2z'^2+4\bar{\Psi}_3z'^3
-\bar{\Psi}_4z'^4=0 \label{eq:tor'}
\ee
for $z$ and $z'$ have only distinct roots. Then, the relations 
(\ref{eq:torus})-(\ref{eq:torus'}), as being fourth order in 
the parametrs $z$ and $z'$, describe a 2-dimensional torus 
\footnote{This is a well known fact of classical algebraic geometry, see for 
example \cite{bi:Mum}. 
I am very grateful to Roger Penrose for clarifying this to me 
\cite{bi:pepr}.}.\\
Let us comment on ({\bf a}) and ({\bf b}).\\
({\bf a}) We know that $w$ and $w'$ can not be zero by their definitions.  
So to have a torus fibration over the Einstein spacetime we need to accept 
that $w$ and $w'$ may vanish. A price paid for this is that some 
of the forms $(F,\bar{F},T,\La,E,\bar{E},\Om,\bar{\Om},\G,\bar{\G})$ will be 
singular on the resulting fibration at these values of $w$ and 
$w'$.\footnote{At the 
worst case the singularity will be a branch point of the fourth order in the 
variables $z$ or $z'$.} With these remarks, from now on, we accept that 
$w$ and $w'$ may vanish. This enables us for the introduction of 
the fibration $\pn$ over $\cal M$ which, over the neighbourhood 
${\cal O}\subset\cal M$ is  
given by
\be
\pn=\pn_1\cup\pn_2,
\ee
where 
\be
\pn_1=\{~(x;w,z,y)\in({\cal O}\times{\bf C}^3)~~~{\rm s.t.}~~~
\Psi y+\frac{1}{4}\Psi_z=0, ~~~\Psi+w^2=0~\},
\ee
\be
\pn_2=
\{~(x;w',z',y')\in({\cal O}\times{\bf C}^3)~~~{\rm s.t.}~~~
\Psi' y'+\frac{1}{4}\Psi'_{z'}=0,~~~ \Psi'+w'^2=0~\}
\ee
and the transition functions between $(w,z,y)$ and $(w',z',y')$ 
coordinates are given by (\ref{eq:wzy}).\\
\noindent
({\bf b}) The discussion in ({\bf a}) means that $\pn$ 
is a torus fibration over $\cal O$ 
provided that the equations (\ref{eq:tor})-(\ref{eq:tor'}) have distinct 
roots. It is well known that the number of distinct roots in 
(\ref{eq:tor})-(\ref{eq:tor'}) is directly related to the algebraic 
(Cartan-Petrov-Penrose \cite{bi:co,bi:pen1,bi:pet}) classification of 
spacetimes. Thus, if the spacetime $\cal M$ is algebraically general in 
$\cal O$ then 
$\pn$ is a torus fibration over $\cal O$. In the algebraically special cases 
the fibres of $\pn$ are degnerate tori. These topologically are:
\begin{itemize}
\item[(II)] a torus with one vanishing cycle in 
the Cartan-Petrov-Penrose type II,
\item[(D)] two sphers touching each other in two different points 
in the Cartan-Petrov-Penrose type D,
\item[(III)] a sphere with one singular point 
in the Cartan-Petrov-Penrose type III,
\item[(N)] two spheres touching each other in one point 
in the Cartan-Petrov-Penrose type N.
\end{itemize}
The pure situations considered so far may be a bit more complicated when 
the Cartan-Petrov-Penrose type of the Einstein spacetime vary from point 
to point. 
Imagine, for example, that along a continuous path 
from $x$ to $x'$ in $\cal O$ the 
Cartan-Petrov-Penrose type of the Einstein spacetime changes from I to II. 
Then  
the fibre of $\pn$ over $x'$ is only a torus with one vanishing cycle 
although the fibre over the starting point $x$ was a torus. It is clear that 
more complicated situations may occur, and that the fibres of $\pn$ 
over different points of $\cal M$ can have topologies II, D, III and N. 
Fibrations of this type are widely used in algebraic geometry. They are 
called elliptic, since their fibres can be any kind (even degenerate) of an 
elliptic curve.\\

\noindent
{\large {\bf 7.}} In Sections 4-6, for the clarity of presentation, we 
restricted ourselves 
to the neighbourhood $\cal O$ of $\cal M$. We ended up with an 
elliptic fibration $\pn$ over $\cal O$. This, however, can be easily 
prolonged to an elliptic fibration over 
the whole $\cal M$. To see this it is enough to observe that  
a fibre over any point in $\cal O$ 
is essentially defined by the Weyl tensor. Since the Weyl tensor is uniquely 
defined on the whole $\cal M$ then we can use equations like 
(\ref{eq:torus}) to uniquely define the elliptic fibres over all $\cal M$. \\

\noindent
Summing up the information from sections 4, 5, 6 we have the following 
theorem.
\newpage
\bt
Given a 4-dimensional spacetime $\cal M$ satisfying the Einstein equations 
$R_{ij}=\la g_{ij}$ one naturally defines a fibration $\Pi:\pn\to\cal M$ 
with the following properties.
\begin{itemize}
\item[(1)] A fibre $\Pi^{-1}(x)$ over a point $x\in\cal M$ is a 
(possibly degenerate) elliptic curve $\cal C$ given by 
$${\cal C}={\cal C}_1\cup{\cal C}_2,$$
$${\cal C}_1=\{~(w,z)\in{\bf C}^2~~~ {\rm s.t.}~~~ w^2=
-\bar{\Psi}_4+4\bar{\Psi}_3z-6\bar{\Psi}_2z^2+4\bar{\Psi}_1z^3
-\bar{\Psi}_0z^4~\},$$ 
$${\cal C}_2=\{~(w',z')\in{\bf C}^2~~~ {\rm s.t.} ~~~
w'^2=
-\bar{\Psi}_0+4\bar{\Psi}_1z'-6\bar{\Psi}_2z'^2+4\bar{\Psi}_3z'^3
-\bar{\Psi}_4z'^4~\},$$ 
where the transition functions between $(w,z)$ and $(w',z')$ 
coordinates are given by  
$$
w=-\frac{w'}{z'^2}\quad\quad\quad z=\frac{1}{z'}. 
$$
\item[(2)] The degeneracy of a fibre depends on the algebraic type 
of the spacetime metric and may change from point to point.
\item[(3)] There is a unique construction of a certain surface $\Pn$ 
of dimension six immersed in the bundle of null coframes $\fm$. $\Pn$ is 
fibred over $\cal M$ and $\pn$ may be viewed as an extension of $\Pn$ 
achieved by adding to each fibre of $\Pn$ at most four points. 
\item[(4)] There are ten 1-forms  
$(F,\bar{F},T,\La,E,\bar{E},\Om,\bar{\Om},\G,\bar{\G})$ on $\pn$ 
with the following properties.
\begin{itemize}
\item[(a)] Forms $T$, $\La$ are real, all the other are complex valued.
\item[(b)] The forms are defined in two steps. First, by  
restricting the soldering 
form components $e^i$ and the Levi-Civita connection components $\om^{i~}_j$ 
from $\fm$ to $\Pn$ and second, by extending the restrictions to $\pn$.
\item[(c)] $(F,\bar{F},T,\La,E,\bar{E})$ constitute the basis of 1-forms on 
$\pn$.
\item[(d)] The forms satisfy the following equations on $\pn$. 
\beq
&{\rm d}F=(\Om -\bar{\Om})\dz F+E\dz T+\G\dz\La,\nonumber\\
&{\rm d}T=\bar{\G}\dz F+\G\dz \bar{F}-(\Om +\bar{\Om})\dz T,\nonumber\\
&{\rm d}\La=\bar{E}\dz F+E\dz\bar{F}+ (\Omega+\bar{\Omega})\dz\La,\\
&{\rm d}E=2\Om\dz E+\bar{F}\dz T+\al F\dz\La,\nonumber
\eeq
with a certain function $\al$ on $\pn$.
\end{itemize}
\end{itemize}
\et
The explicit formulae for 
the forms on $\pn_1$ (resp. on $\pn_2$) 
may be obtained from the expressions of Lemma 1 (resp. Lemma 2) by inserting 
the relations $y=-\Psi_z/(4\Psi)$ and $w^2+\Psi=0$ (resp. 
$y'=-\Psi'_{z'}/(4\Psi')$ and $w'^2+\Psi'=0$).\\ 

\noindent
{\large {\bf 8.}} Finally we note that the bundle $\pn$ constitutes a double branched 
cover of the Penrose bundle $\Pen$ of null directions over the spacetime.\\ 
To see this consider the map $f$ given by
$$\pn_1\ni (x;w,z,y)\stackrel{f}{\to}(x,z)\in {\cal O}\times{\bf C}$$
$$\pn_2\ni (x;w',z',y')\stackrel{f}{\to}(x,z')\in {\cal O}\times{\bf C}.$$
Since on the intersection $\pn_1\cap\pn_2$ the coordinates 
$z$ and $z'$ are related by $z'=1/z$, then the two copies of {\bf C}, 
which appear in the above relations may be considered as two coordinate 
charts (say around the North and the South pole, respectively) on the 
2-dimensional sphere. This sphere is the sphere of null directions at 
a given point of $\cal O$ which can be seen as follows. \\
Consider directions of all the 1-forms $L(z)=L+z\bar{z}K+z\bar{M}+\bar{z}M$
at $x\in\cal O$ for all the values of the complex paramter $z$. These 
directions are in one to one coorespondence with 
null directions $k(z)=k+z\bar{z}l-zm-\bar{z}\bar{m}$ via $L(z)=-g(k(z))$. 
These null directions do not form a sphere yet, since the direction 
corresponding to the vector $l$ is missing. But the missing direction is 
included in the 
family of null directions corresponding to the directions of 
1-forms $K(z')=K+z'\bar{z}'L+z'\bar{M}+\bar{z}'M$ at $x$. Thus the 
space parametrized by $z$ and $z'$ subject to the relation $z'=1/z$  
is in one to one correspondence with the sphere of null directions at $x$. 
This proves the assertion that $f$ is a cover of $\Pen$.\\ 
The map $f$
is a double cover since if $z$ (or $z'$) is not a root of 
(\ref{eq:tor}) 
(resp. (\ref{eq:tor'})) then $f^{-1}(z)$ is a two-point set. In at most  
four cases when $z$ is a root $f^{-1}(z)$ is a 
one-point set. This proves that $f$ is a singular cover..\\ 
As we 
know the singular points are branch points, which proves the statement 
from the begining of this section.\\

\noindent
{\Large Acknowledgements}\\
I am deeply indebted to Andrzej Trautman who foccused my attention on 
the problem of encoding Einstein equations on natural bundles associated 
with a spacetime. Without his influence this work would never have been 
completed.\\ 
I would also like to express my gratitude to Paolo Budinich, 
Mike Crampin, Ludwik D\c{a}browski, Jerzy Lewandowski, Roger Penrose, 
David Robinson and Jacek Tafel for helping me at various stages of the 
present work. 

\newpage
\end{document}